\documentclass[
  journal=pasa,
  manuscript=research-paper, 
  year=2020,
  volume=37,
]{cup-journal}

\usepackage{microtype,siunitx,booktabs}
\usepackage{amssymb}
\title{Evidence of distant spiral arms in the Galactic disk quadrant IV from VVV red clump giants}

\author{R. Kammers}
\affiliation{Departamento de  F\'{i}sica, Universidade  Federal de
  Santa Catarina, Trindade 88040-900, Florian\'opolis, SC, Brazil}
\email[R.~Kammers]{betokammers@gmail.com}

\author{R. K. Saito}
\affiliation{Departamento de  F\'{i}sica, Universidade  Federal de
  Santa Catarina, Trindade 88040-900, Florian\'opolis, SC, Brazil}

\author{E. Botan}
\affiliation{Instituto de Ci\^{e}ncias Naturais, Humanas e Sociais, Universidade Federal de Mato Grosso, Res. Cidade Jardim, 78550-728, Sinop, MT, Brazil}
\alsoaffiliation{Departamento de  F\'{i}sica, Universidade  Federal de
  Santa Catarina, Trindade 88040-900, Florian\'opolis, SC, Brazil}

\author{D. Minniti}
\affiliation{Instituto de Astrofísica,  Facultad  de  Ciencias  Exactas, Universidad  Andres Bello,  Av.  Fernandez  Concha 700,  Las Condes, Santiago, Chile}
\alsoaffiliation{Vatican Observatory, V00120 Vatican City State, Italy}

\author{J. Alonso-Garc\'{i}a}
\affiliation{Centro  de Astronom\'{i}a  (CITEVA),  Universidad de  Antofagasta,
  Av. Angamos 601, Antofagasta, Chile}
\alsoaffiliation{Millennium Institute of Astrophysics, Nuncio Monse\~nor Sotero Sanz 100, Of. 104, Providencia, Santiago, Chile}

\author{L. C. Smith}
\affiliation{Institute of Astronomy, University of Cambridge, Madingley Road,
  Cambridge, CB3 0HA, UK}
  
\author{P. W. Lucas}
\affiliation{Centre for  Astrophysics Research, School of  Physics, Astronomy
  and Mathematics, University of Hertfordshire, College Lane,\\ 
  Hatfield AL10 9AB, UK}  

\doi{10.1017/pasa.2020.32}

\received {dd Mmm YYYY}
\revised  {dd Mmm YYYY}
\accepted {dd Mmm YYYY}
\published{22 September 2020}


\begin{document}

\begin{abstract}
The discovery of new clear windows in the Galactic plane using the VVV
near-IR extinction maps allows the study of the structure of the Milky
Way (MW)  disk. The ultimate  goal of this work  is to map  the spiral
arms  in the  far side  of the  MW, which  is a  relatively unexplored
region  of  our  Galaxy,  using  red clump  (RC)  giants  as  distance
indicators.   We  search for  near-IR  clear  windows located  at  low
Galactic   latitudes ($|b|< 1$\ deg)   in the   MW disk  using the  VVV
near-IR     extinction    maps.  We   have  identified two new windows
named VVV\ WIN\ 1607$-$5258 and  VVV\ WIN\ 1475$-$5877, respectively,
that complement the previously known  window  VVV\ WIN\ 1713$-$3939. We
analyse  the distribution  of RC  stars in  these three  clear near-IR
windows  and measure  their number  density along  the line  of sight.
This allows us  to find overdensities in the  distribution and measure
their distances  along the line of  sight. We then use  the VVV proper
motions  in  order to  measure  the  kinematics  of  the RC  stars  at
different   distances.   We   find   enhancements   in  the   distance
distribution of  RC giants  in all  the studied  windows, interpreting
them as  the presence of  spiral arms in the  MW disk. These structures
are absent in the current models of synthetic population for the same
MW lines of sight. We were able to trace the end of  the Galactic  bar,
the  Norma arm,  as well  as the Scutum Centaurus arm  in the far disk.
Using the  VVV proper motions, we measure the kinematics for these
Galactic features, confirming that they share the bulk rotation of the
Galactic disk.
\end{abstract}

\section{Introduction}

The Milky  Way (MW) is a  barred spiral galaxy suggested  to present a
4-arms  pattern.  The  MW spiral  arms  have long  been recognized  as
presenting  coherent arcs  and  loops  in Galactic  longitude-velocity
$(V_l)$     plots     of     atomic    and     molecular     emissions
\citep[e.g.][]{2014ApJ...783..130R,2017AstRv..13..113V}.    The  exact
location  and detailed  structure of  the  MW spiral  arms are  highly
uncertain, and part of the current  models to the outer MW regions are
extrapolations    of   the    observed   in    the   closer    regions
\citep[e.g.][]{2015MNRAS.450.4150C,2016ARA&A..54..529B,2019ApJ...885..131R}.
Due to effects  caused by warping, the outer disk  slowly departs from
the Galactic Plane  in both HI and stars. In  addition to the warping,
the  outer disk  flares and  possesses a  fair degree  of substructure
\citep{2016ARA&A..54..529B}.

\begin{figure*}
   \centering 
   \includegraphics[scale={1.3}]{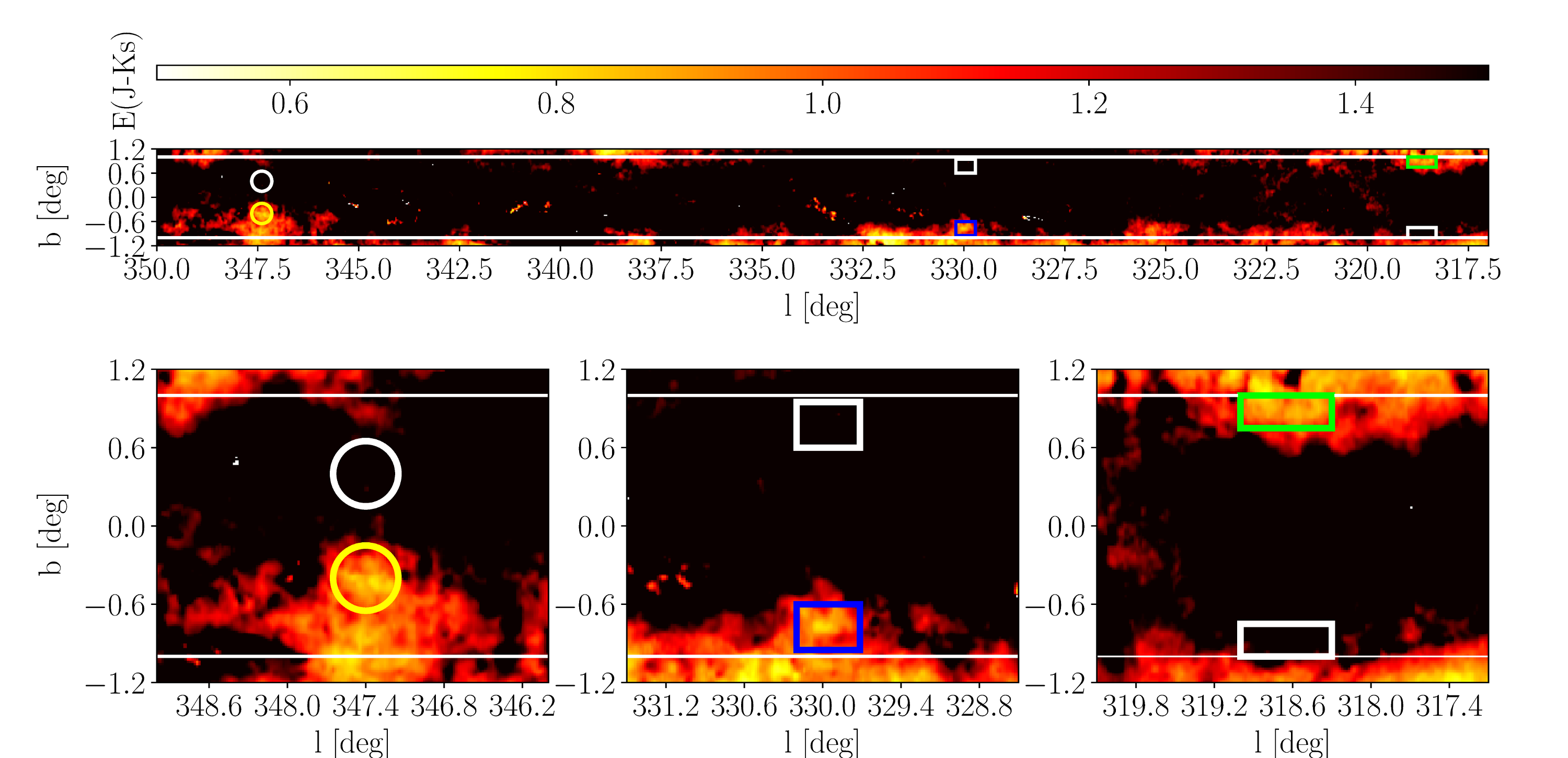}
   \caption{Top panel:  a modified version  of the VVV  extinction map
     for     the      southern     Galactic      plane     \citep[from
     ][]{2018A&A...616A..26M}  with   all  regions   with  $E(J-K_{\rm
       s})>$1.5  masked  in black  color  in  order to  highlight  low
     extinction regions.   Horizontal white  lines mark the  region of
     interest ($-1.0$ deg $< b <$  $1.0$ deg). In this map the
       previously  known low  extinction window  VVV WIN  1713–3939 is
       marked  as a  yellow circle  while  the two  newly found  windows,
       VVV~WIN~1607$-$5258  and VVV~WIN~1475$-$5877, appear as  a blue
       and  a  green  rectangle,  respectively.  For  each  window  we
       selected  also a  control  field with  same  area at  symmetric
       latitudes,  marked with  a white  circle or  rectangle. Bottom
     panels: a  zoomed view  around the region  of VVV~WIN~1713$-$3939
     (left   panel),    VVV~WIN~1607$-$5258   (central    panel)   and
     VVV~WIN~1475$-$5877 (right panel).}
\label{fig:map}
\end{figure*}

The high extinction  towards the Galactic plane is  the major obstacle
to prevent  the study  of the  far side  of the  MW.  To  minimize the
problem, windows of  low extinction such as the Baade's  window in the
Galactic                                                         bulge
\citep{1951POMic..10....7B,1965ApJ...141...43A,1971AJ.....76.1082V}
have  been used  in  the study  of the  stellar  content and  Galactic
structure,     among    others.      \cite{2009PASP..121..213C}    and
\cite{2012A&A...545A..39R} show that only Scutum-Centaurus and Perseus
can  be considered  main arms,  associated with  overdensities of  old
stellar disk  and OB stars,  while Norma and Sagittarius  arms present
younger stars and a few OB stars.

The ESO VISTA Variables in  V\'{\i}a L\'actea (VVV) near-IR survey has
recently scanned a total of 562~sq.~deg.  area of the MW bulge and the
adjacent plane. The VVV strategy  consisted in two set of observations
in  $ZYJHK_{\rm s}$  filters,  plus $50-100$  observations in  $K_{\rm
  s}$~band over many years,  providing a deep, high-resolution dataset
of       the       inner       Galaxy       in       the       near-IR
\citep{2010NewA...15..433M,2012A&A...537A.107S}.    Using  VVV   data,
\cite{2018A&A...616A..26M} and  \cite{2020MNRAS.494L..32S} reported on
the discovery of two windows of low extinction located right in the MW
plane,    namely    VVV~WIN~1713$-$3939    and    VVV~WIN~1733$-$3349.
VVV~WIN~1733$-$3349  lies  towards  the  inner MW  bulge  at  Galactic
coordinates  ($l,~b=354.8, -0.3$  deg),  while VVV~WIN~1713$-$3939  is
about 7.5 deg  apart in longitude, at  coordinates ($l,~b=347.4, -0.4$
deg).

VVV~WIN~1713$-$3939  is roughly  circular in  shape, with  $30'$
diameter in  size.  The  distribution of red  clump (RC)  stars within
VVV~WIN~1713$-$3939  shows   a  bimodal  distribution   in  magnitude,
interpreted  as the  signature of  the  structure of  the spiral  arms
across the background disk.  The RC  peaks at $d=11.2$ and $15.2$ kpc,
in agreement  with the expected  position for the the  Sagittarius arm
and the end of the long bar,  and the far side of the Scutum-Centaurus
arm, respectively \citep[see Fig. 6 in][]{2018A&A...616A..26M}.

Here  we  complement  the  analysis  of VVV~WIN~1713$-$3939  made  by
\cite{2018A&A...616A..26M} and describe the  search for new windows of
low extinction in  the MW plane, reporting the  discovery and analysis
of VVV~WIN~1607$-$5258  and VVV~WIN~1475$-$5877.  The  RC distribution
and  kinematics   within  these  three   windows  were  used   to  map
overdensities  on the  far  side  of the  disk.   Our  results are  in
agreement with expectations  for the spiral arms both  in distance and
in velocity along the Galactic longitude.

  \section{Two new windows of low extinction}

We  made use  of  the  $E(J-K_{\rm s})$  extinction  map presented  in
\cite{2018A&A...616A..26M}  to search  for windows  of low  extinction
located at  low Galactic latitudes in  the plane of the  MW.  To avoid
departing from the  Galactic plane at larger distances  we limited our
search to  $|b|<1.0$~deg and $310 <l<350$  deg, and then we  applied a
threshold limit  of $E(J-K_{\rm s})$  = 1.5 mag  in order to  mask the
regions of high  extinction.  In the resulting map (see  the top panel
of Fig.~\ref{fig:map}) two more  promising low extinction windows were
detected,  in  addition  to   the  already  known  VVV~WIN~1713$-$3939
\citep{2018A&A...616A..26M}. We  also verified  that these  regions of
low    extinction     are    also     seen    in    the     maps    of
\cite{2006A&A...453..635M},       \cite{2012ApJS..201...35N}       and
\cite{2019MNRAS.488.2650S}, and  in the  spatial distribution  of Gaia
sources \citep{2016A&A...595A...2G}.

In  selecting  RC  stars  from VVV~WIN~1713$-$3939  we  use  the  same
definition      as      in     \citet{2018A&A...616A..26M},      where
VVV~WIN~1713$-$3939 is a  circular area of $30'$  diameter centered at
$(l, b) = (347.4, -0.4)$~deg.

VVV~WIN~1607$-$5258 is  a rectangular  region of $30'$  $\times$ $21'$
size, centered  at $(l, b)  = (329.95,-0.77)$~deg.  For a  latitude of
$b=-0.77$~deg the vertical projection along the line of sight is small
$z < 200$ pc even at the far side  of the MW disk ($d < 15$ kpc).  The
mean extinction  within the  window is $E(J-K_{\rm  s}) =  1.01$ $\pm$
0.16~mag, corresponding to $A_{K_{\rm s}} = 0.49$~mag using $A_{K_{\rm
    s}}/E(J-K_{\rm  s})  =   0.484$  \citep{2018A&A...616A..26M}.   We
define  VVV~WIN~1475$-$5877  also  as   a  rectangular  region,  $42'$
$\times$ $15'$ size,  centered at $(l, b)  = (318.65,+0.87)$~deg.  The
mean reddening  within this  window is $E(J-K_{\rm  s}) =  0.91$ $\pm$
0.06~mag   and   $A_{K_{\rm   s}}   =   0.44$~mag   using   $A_{K_{\rm
    s}}/E(J-K_{\rm  s}) =  0.484$.  Other  low extinction  regions are
also   seen  in   the   modified  $E(J-K_{\rm   s})$   map,  near   to
VVV~WIN~1607$-$5258  ($l=332$~deg)  and around  $l=325.5$~deg.   These
regions are  smaller and more  inhomogeneous in relation to  the three
selected windows and therefore they are not subject of this work.

\begin{figure*}
   \centering \includegraphics[scale={1.0}]{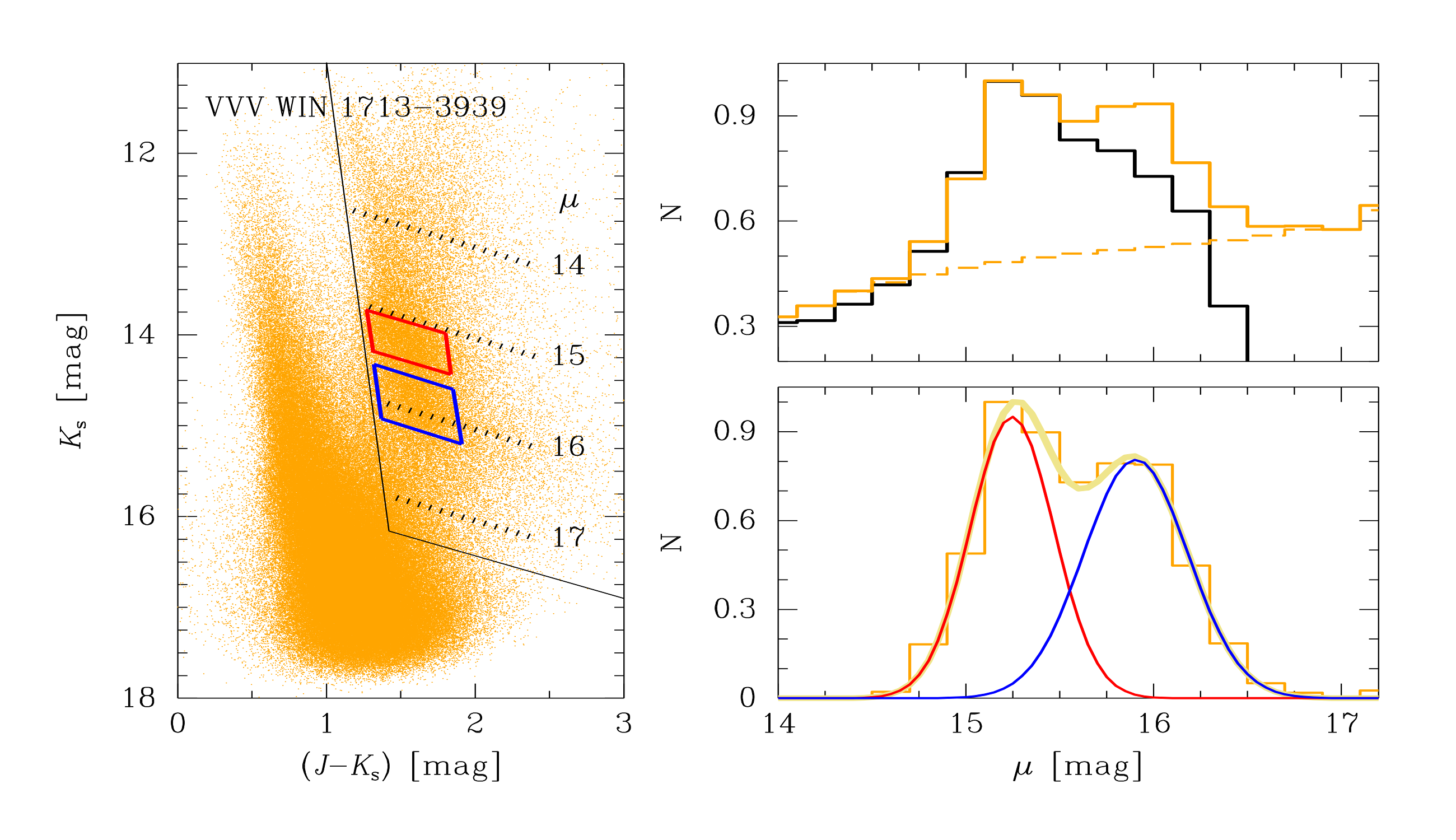}
   \vspace{0cm}
   \caption{Analysis of the RC distances in VVV~WIN~1713$-$3939.  Left
     panel: $K_{\rm s} ~vs~ (J-K_{\rm  s})$ CMD. The region defined as
     the RC  locus is  contoured by  a solid  black line  while dotted
     lines are  the scale  of the  distance modulus  $\mu$ for  the RC
     stars (see Section 2). We note the presence of a double RC in the
     region  around  [$(J-K_{\rm s}),K_{\rm  s}$]$\sim  1.7,14.3$~mag. 
     Diamond shaped boxes mark the regions used for the proper 
     motion measurements (see Section 4 and Fig.~\ref{fig:pm}).
     Top right panel: Distribution of  RC stars versus distance moduli
     within  VVV~WIN~1713$-$3939 (orange)  compared  with its  control
     field  (black). A  farther  overdensity is  clearly  seen in  the
     window  while is  absent  in the  control  field, suggesting  the
     presence  of a  structure  located at  larger distances.   Bottom
     right panel: Multi Gaussian fit for the distribution of window RC
     stars  after  subtracting  a  polynomial fit  to  the  luminosity
     function (yellow dashed line in the top right panel). Information
     about the two peaks are listed in Table 1.}
\label{fig:cmd}
\end{figure*}

\begin{figure*}
   \centering \includegraphics[scale={1.0}]{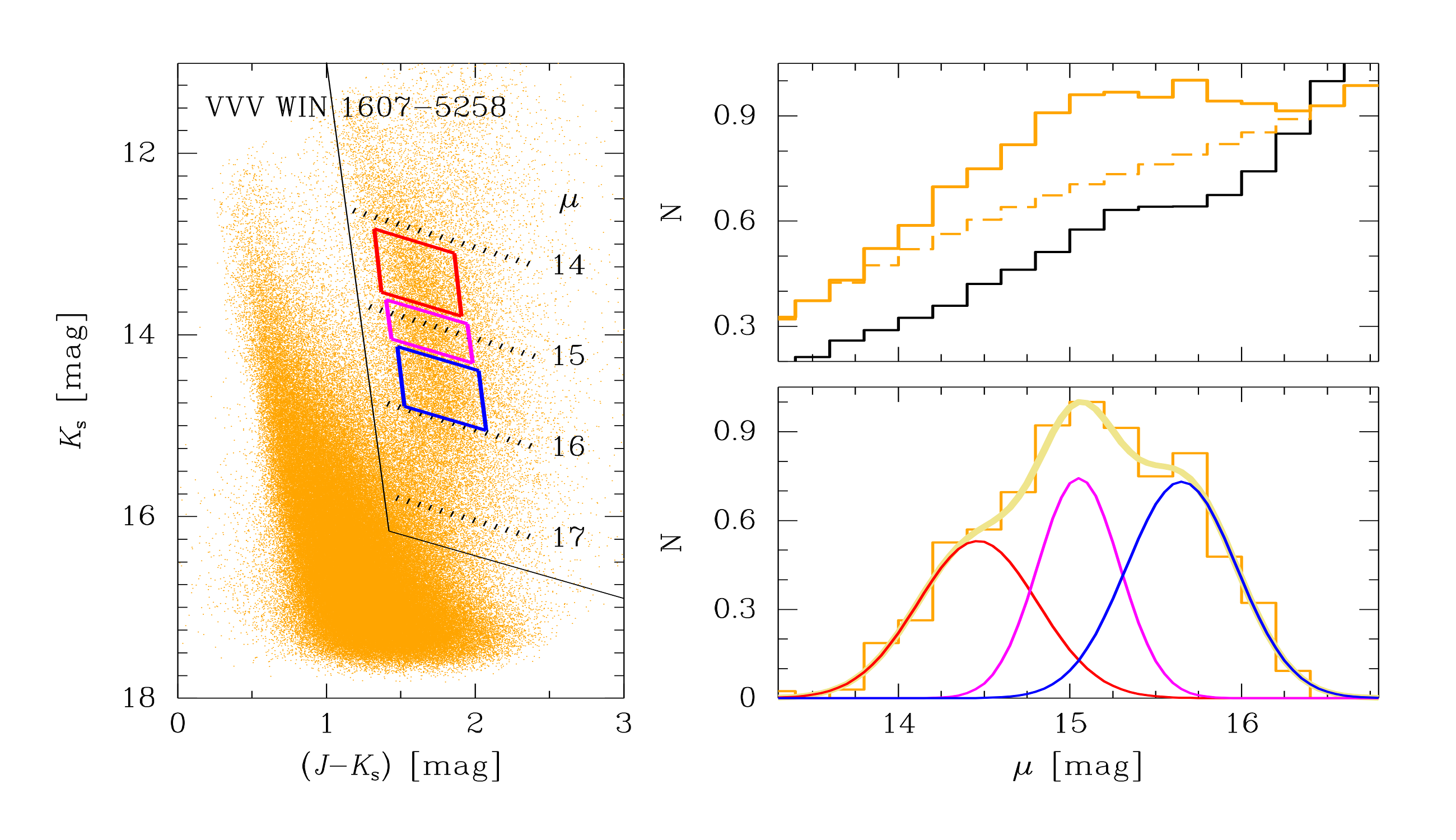}
   \vspace{0cm}
   \caption{Analysis of the RC  distances in VVV~WIN~1607$-$5258.  The
     notation is similiar to that presented in Fig.~2.}
\label{fig:cmd2}
\end{figure*}

\begin{figure*}
   \centering \includegraphics[scale={1.0}]{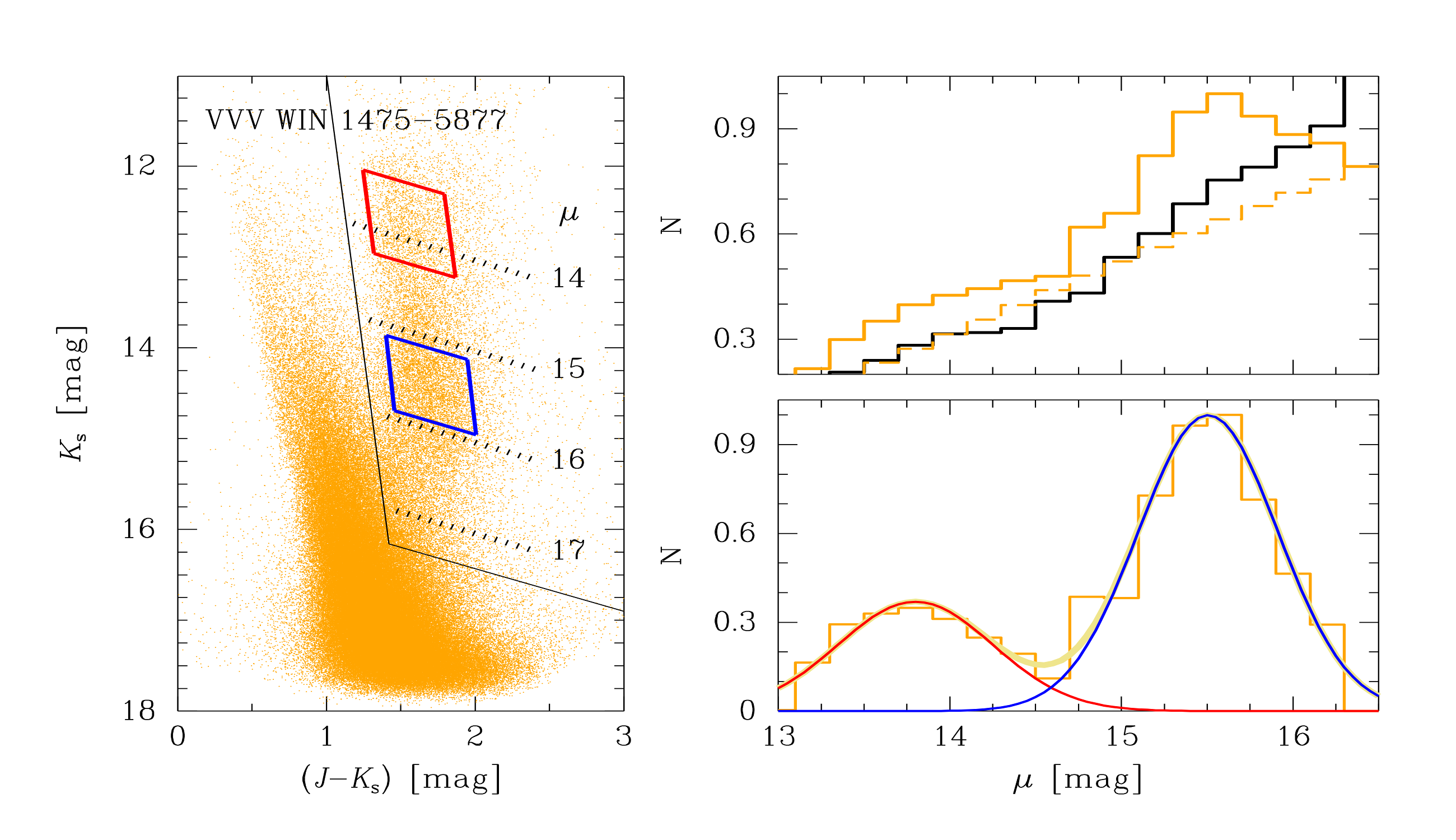}
   \vspace{0cm}
   \caption{Analysis of the RC  distances in VVV~WIN~1475$-$5877.  The
     notation is similiar to that presented in Fig.~2.}
\label{fig:cmd3}
\end{figure*}

Our analysis  of the stellar content  along the line of  sight for the
windows  was  performed  using  red   clump  (RC)  stars  as  distance
indicators. These stars  have been used widely as  standard candles to
trace  the Galactic  structure. A  comprehensive review  about the  RC
stars is  provided by \cite{2016ARA&A..54...95G}. Deeper  VVV PSF data
\citep{2018A&A...619A...4A}   were  used   to  build   color-magnitude
diagrams (CMDs) for the windows,  including an independent analysis of 
  VVV~WIN~1713$-$3939,  updated with  the  new photometry. The
  $K_{\rm  s}  ~vs  ~(J-K_{\rm   s})$  CMDs  for  VVV~WIN~1713$-$3939,
  VVV~WIN~1607$-$5258 and  VVV WIN 1475$-$5877 are  shown in  the left
  panels  of Figs.  \ref{fig:cmd}, \ref{fig:cmd2}  and \ref{fig:cmd3},
  respectively. These figures also summarize  in the right panels  the 
  results on the distances of  the RC  stars as  described below.

The selection of  the RC region was made case  by case using a
  color cut  directly on the  CMDs. The selection  made to separate  the red
  giant stars from the stars of the main sequence does not need a very
  precise cut, since  the stars of the RC are  the most numerous stars
  in the giant branch. The distribution  of the RC distances was then
calculated  using  the  distance  modulus,   $\mu  =  -5  +  5\,log\,d
(pc)=K_{\rm s}-0.484\times(J- K_{\rm s})+1.924$. This equation assumes
the   RC   absolute  magnitudes   calibrated   using   Gaia  data   by
\citet{2018A&A...609A.116R}, where  $M_{K_{\rm s}}= -1.605  \pm 0.009$
and  $(J-K_{\rm s})  =  0.66  \pm 0.02$  mag,  as  well as  $A_{K_{\rm
    s}}/E(J-        K_{\rm       s})        =       0.484$        from
\cite{2018A&A...616A..26M}.  The   distribution  in  distance
  modulus for the three windows is  presented top right panels of
  Figs.           \ref{fig:cmd},          \ref{fig:cmd2}           and
  \ref{fig:cmd3}.  \citet{2011AJ....142...76S}  argue  that  not  all
selected stars  are RCs, but the  presence of subgiant branch  and RGB
stars, or  even main-sequence stars at  fainter magnitudes, contribute
to the  smooth underlying background.  These  different populations do
not  affect   the  location  of   any  sharp  edges  of   the  stellar
distribution, which  can, however, be well  defined by the RC.  We can
assume the same behavior for the entire disk.

We applied  then a polynomial  fit to the  luminosity function
  (LF)  and subtracted  it  from  the LF,  aiming  to  obtain the  net
  distribution of RC stars, free of the background LF sources (see the
  bottom     right    panels     of     Figs.     \ref{fig:cmd}     to
  \ref{fig:cmd3}). Finally, a  multi Gaussian fit was  applied for the
  distribution of window RC stars, and  the mean distances for the RCs
  were found.  This technique  is widely used  in studies  of galactic
  structure  \citep[e.g.,][]{2007MNRAS.378.1064R, 2011ApJ...733L..43M,
    2015MNRAS.450.4050W,  2018MNRAS.481L.130G,  2019A&A...623A.168S}.
The  peak  of  the  distribution adjusted  through  the  Gaussian  fit
represents the  position of the highest  density of RC stars  for this
distribution,  while the  errors  estimates are  obtained through  the
standard  deviation of  the  sample data. The distances  and
  respective uncertainties calculated for the  RC peaks for the three
  windows are shown in Table~1.

For  the three  windows  the  RC distribution  in  distance shows  the
presence of distant secondary peaks at larger distances.  The presence
of  a  faint  RC  peak  in  Galactic  CMDs  is  sometimes  subject  of
controversy since it  has also been interpreted as  the red-giant bump
\citep[e.g.][]{2011ApJ...730..118N,2013ApJ...766...77N} instead of the
signature of a secondary, farther structure  in the same line of sight
\citep[e.g.][]{2011A&A...534L..14G,2011AJ....142...76S,2018MNRAS.481L.130G}.
To verify  the hypothesis  that our  RCs map  the presence  of farther
structures,  we selected  symmetric  control fields  in latitude  (see
Figs.~\ref{fig:map})  and then  applied  the  same analysis  described
above.  These  fields should  present the  same stellar  population as
within the windows, however embedded in higher levels of dust. Similar
to VVV~WIN~1713$-$3939 \citep{2018A&A...616A..26M}, the secondary peak
is not present in the control fields (see the top right panels
  of  Figs.~\ref{fig:cmd}   to  \ref{fig:cmd3}).  Since   warping  is
negligible at these coordinates \citep[e.g.  ][]{2006A&A...451..515M},
one  can interpret  the  second  peak as  the  presence  of a  distant
structure only seen within the windows where the extinction is low. In
the control fields this structure is not visible because it is located
behind the dust.

\begin{table*}
\centering
\begin{tabular}{lccc}
\hline
\hline
Window Name & RC Peaks          & PM              & Tangencial vel.         \\
            & Distance (kpc)    & (mas\,y$^{-1}$)  & $l$ coord. (km\,s$^{-1}$) \\
\hline
VVV WIN     &  11.2 $\pm$ 2.3 & $-$7.19 $\pm$ 2.28 & $-$381.64 $\pm$ 121.05 \\ 
1713$-$3939 &  15.2 $\pm$ 3.9 & $-$6.09 $\pm$ 2.67 & $-$438.82 $\pm$ 192.39 \\ 
\hline                                      
VVV WIN     &   7.8 $\pm$ 2.5 & $-$6.87 $\pm$ 1.72 & $-$254.02 $\pm$ 63.60 \\ 
1607$-$5258 &  10.2 $\pm$ 2.2 & $-$7.01 $\pm$ 2.01 & $-$338.95 $\pm$ 97.19 \\ 
            &  13.5 $\pm$ 4.0 & $-$6.47 $\pm$ 2.49 & $-$414.06 $\pm$ 159.35 \\ 
\hline                                      
VVV WIN     &   5.8 $\pm$ 2.4 & $-$6.72 $\pm$ 1.50 & $-$184.76 $\pm$  41.24 \\ 
1475$-$5877 &  12.6 $\pm$ 4.7 & $-$6.27 $\pm$ 1.76 & $-$374.51 $\pm$ 105.12 \\ 
\hline
\hline
\end{tabular}
\caption{Physical distances, proper  motions and tangential velocities
  from  RC estimations  in the  three studied  windows. Distances  and
  velocities   are   also   presented  in   Figs.~\ref{fig:view}   and
  ~\ref{fig:vel}, respectively.}
\end{table*}

\section{Model comparisons}

\begin{figure}
   \includegraphics[scale={0.95}]{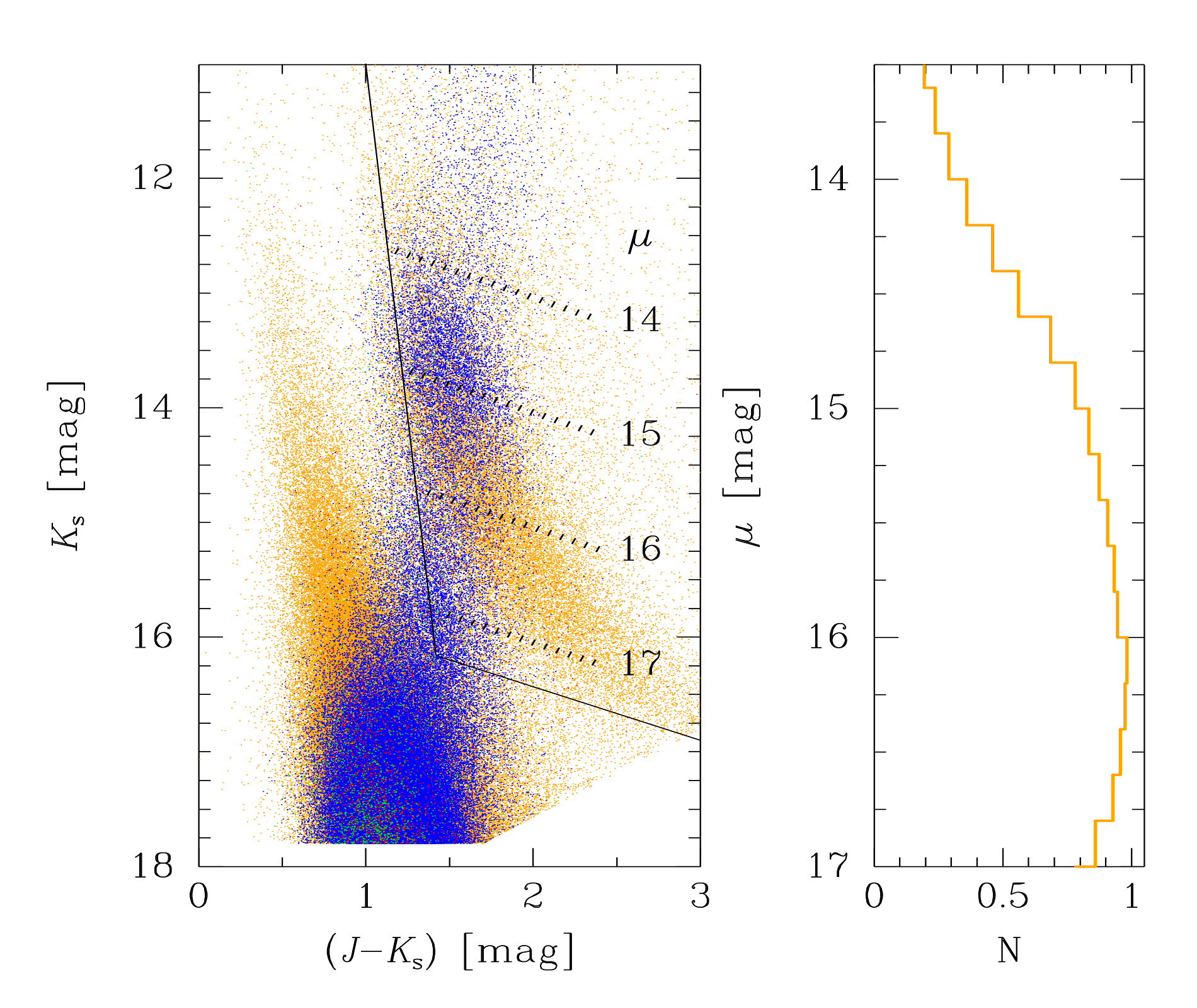}
\caption{Left panel:  Trilegal synthetic  $K_{\rm s}$  $vs$ $(J-K_{\rm
    s})$  CMD for  VVV WIN  1713-3939.  Data points  are colour  coded
  according  with  the Galactic  component:  thin-disk  are in  orange
  colour (63.9\% of  the sources), bulge in  blue (34.6\%), thick-disk
  in   red   (0.9\%)  and   halo   in   green  (0.6\%).   Similar   to
  Figs.\ref{fig:cmd} to  \ref{fig:cmd3}, the region defined  as the RC
  locus is contoured by a solid line while dotted lines mark the scale
  of  the distance  modulus.  Right panel:  distribution  of RC  stars
  versus distance moduli as selected in the corresponding CMD.}
\label{fig:mod01}    
\end{figure}

\begin{figure}
   \includegraphics[scale={0.95}]{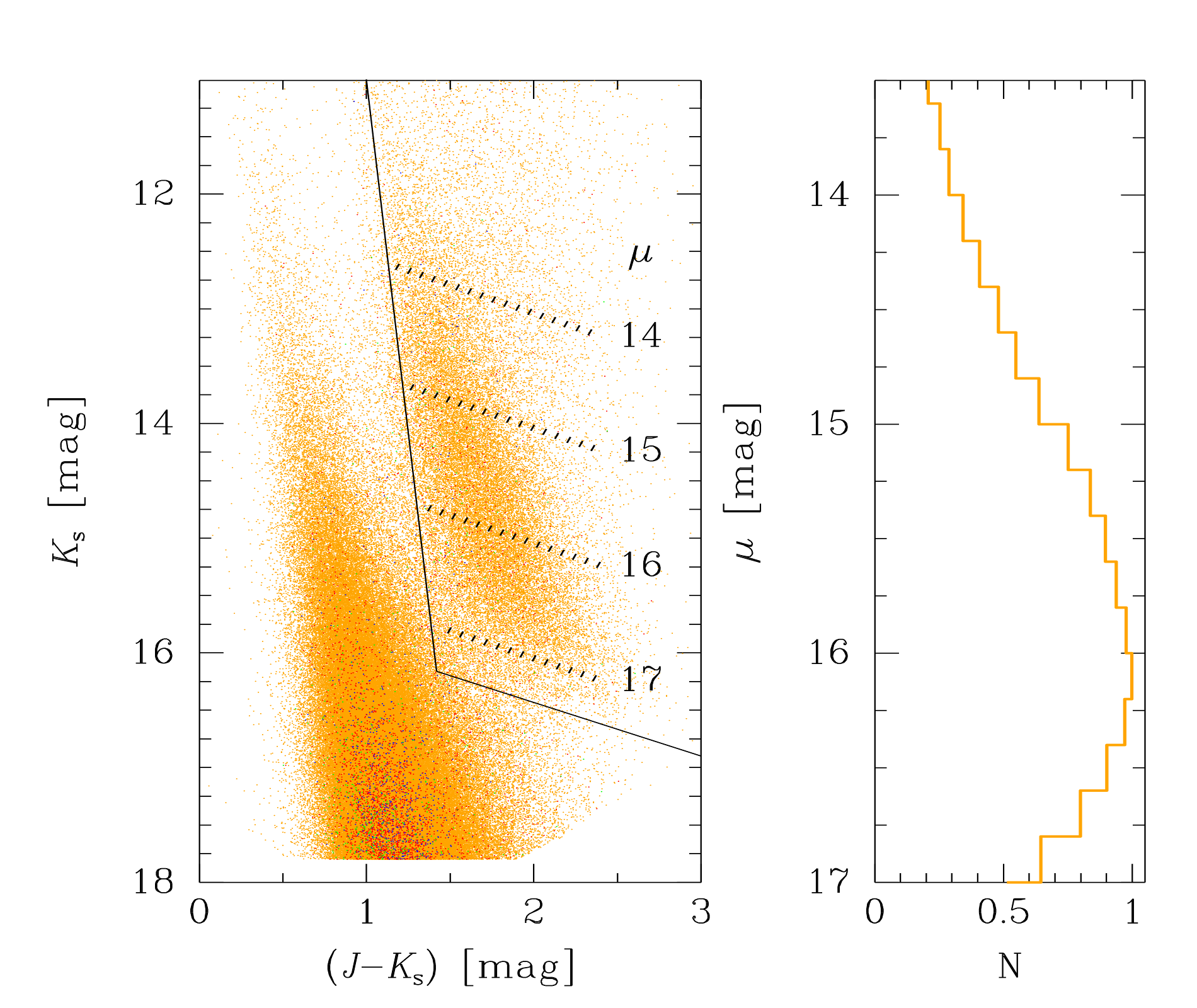}
\caption{Trilegal synthetic  $K_{\rm s}$ $vs$ $(J-K_{\rm  s})$ CMD for
  VVV WIN 1607–5258.  The thin disk (in orange)  corresponds to 98.2\%
  of the sources. The notation is similar to Fig.~\ref{fig:mod01}}
\label{fig:mod02}    
\end{figure}

\begin{figure}
   \includegraphics[scale={0.95}]{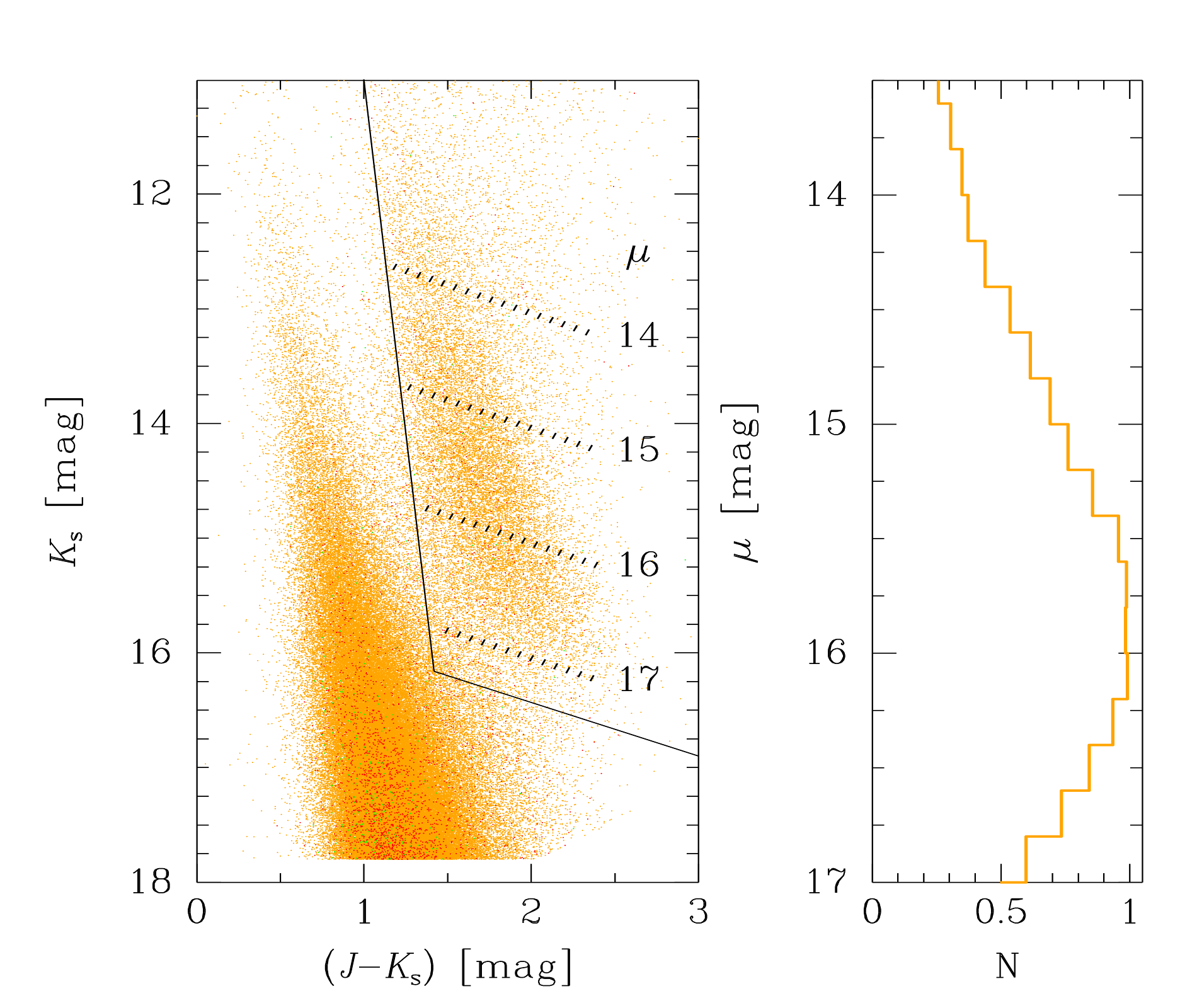}
\caption{Trilegal synthetic  $K_{\rm s}$ $vs$ $(J-K_{\rm  s})$ CMD for
  VVV  WIN  1475$-$5877.  Thin  disk  (in orange)  is  98.0\%  of  the
  sources. The notation is similar to Fig.~\ref{fig:mod01}}
\label{fig:mod03}    
\end{figure}

In  this section  we compare  our  observational data  with a  stellar
population  synthesis model.  In  this  exercise we  made  use of  the
Trilegal  \citep{2005A&A...436..895G}  model  to  analyze  the  region
around the red  clump in the synthetic data. Trilegal  is available in
the VISTA $JHK_{\rm  s}$ colours, and labels the  sources according to
the Galactic  component (bulge, halo,  thin- and thick-disk).  For the
comparison we  selected regions of 0.25  sq deg area, centered  at the
same central coordinates for the three low extinction windows: ($l,b$)
=  ($347.4, -0.4$)~deg  for  VVV WIN  1713–3939,  ($l,b$) =  ($329.95,
-0.77$)~deg for VVV WIN 1607–5258  and ($l,b$) = ($318.65, +0.87$)~deg
for VVV WIN 1475–5877. CMDs for the three regions are presented in the
left    panels   of    Figs.~\ref{fig:mod01},   \ref{fig:mod02}    and
\ref{fig:mod03}.

Using the  Trilegal flags for  the galactic components, we  found that
for VVV WIN  1713-3939 there is a contamination by  bulge sources in a
fraction of 34.6\%,  however the majority of the sources  are from the
thin disk  (63.9\%). The  sum of  halo and  thick disk  corresponds to
1.5\%.  For  the  other  two  windows, thin  disk  sources  are  fully
dominant:  98.0\%   for  WIN   1607–5258  and   98.2\%  for   VVV  WIN
1475–5877. For the selection of the RC  region we use the same cuts as
in  the  observational  CMDs,  as  well  as  the  same  steps  in  the
calculation of the distance modulus and its distribution.

The  distribution of  RC stars  versus distance  moduli for  the three
windows are  presented in  the right panels  of Figs.~\ref{fig:mod01},
\ref{fig:mod02} and  \ref{fig:mod03}. In  all cases  the distributions
are smooth and  seem to follow the luminosity function,  with the star
counts increasing  down to  $K_{\rm s}>16$~mag. At  fainter magnitudes
the counts vanish,  as expected in a distribution along  a finite disk
\citep[e.g.,][]{2011ApJ...733L..43M}. We  note that  in all  cases the
distributions  do not  show  any evidence  of the  overdensities/peaks
related to the RC as presented in the observational data. Since spiral
arms   are    not   modeled   by   the    Trilegal   synthetic   model
\citep{2005A&A...436..895G}, that  reinforces our  interpretation that
the RC overdensities are caused  by structures (spiral arms) along the
line of sight of each window.

\section{Proper motions and the tangential velocities}

\begin{figure}
\includegraphics[scale={1.6}]{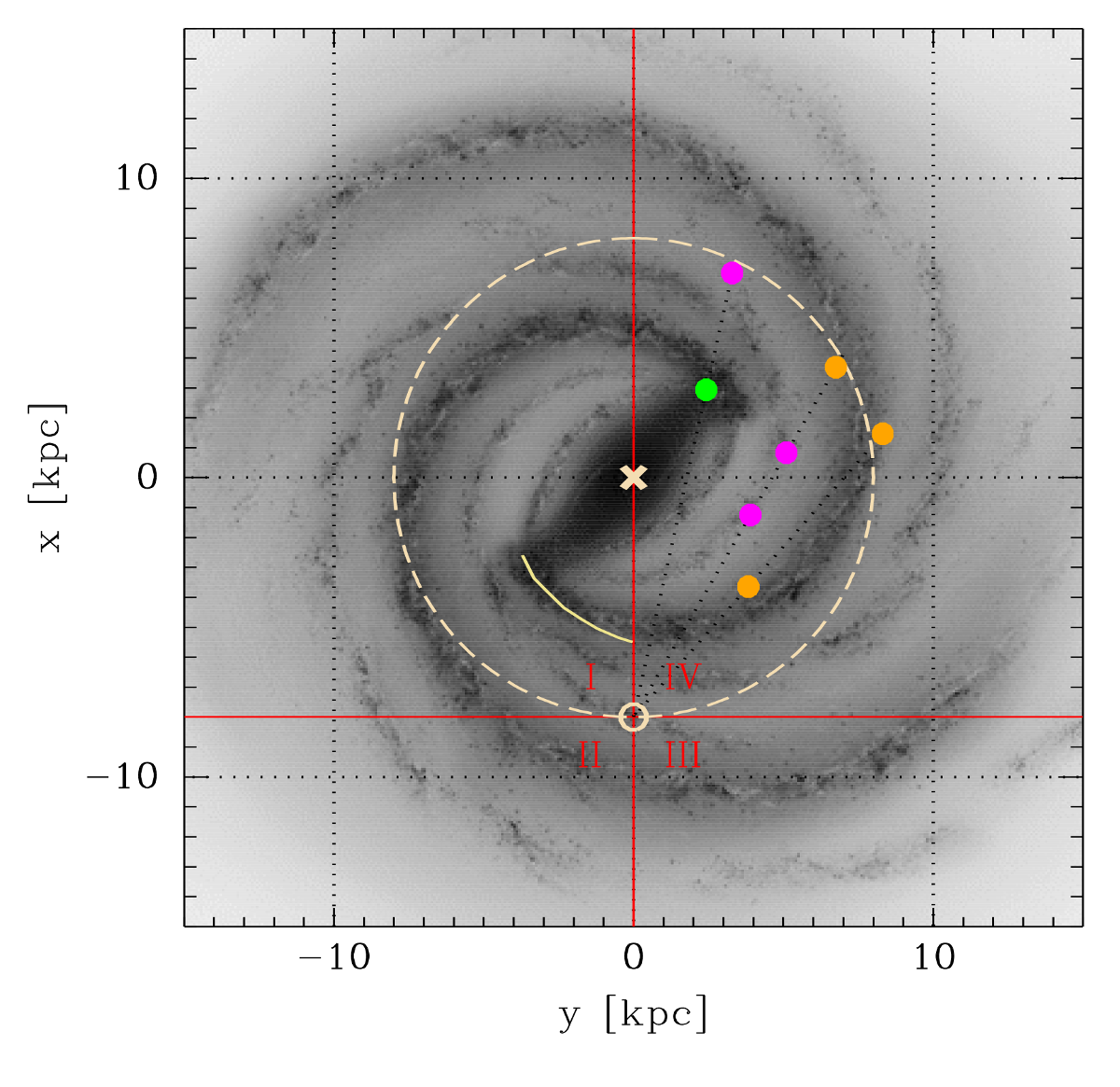}
\caption{Schematic    map   of    the   Milky    Way   adapted    from
  \citet{2009PASP..121..213C}  with the  Galactic quadrants  marked in
  red.  The position  of each RC overdensity along the  lines of sight
  for the  three low extinction windows  are overploted on the  map as
  solid circles.  We adopted  the distance to  the Galactic  centre as
  $R_0$ =  8.18 kpc.  The green  circle shows the Perseus  arm (or the
  end of the  Galactic bar), magenta circles mark the  Norma arm while
  orange circles mark the position of Scutum-Centaurus arm.  The solid
  line traces  the Scutum-Centaurus arm accordingly  with \citet[][see
    Section 4]{2018A&A...618A.168R}.  The same  colour code is used in
  Fig.~\ref{fig:vel}.}
\label{fig:view}
\end{figure}

The  magnitude difference  between the  two RC  peaks changes  in each
window, thus  assuming that the  stellar populations do not  change at
those different  positions, the double peak  cannot be the RC  and the
RGB-bump,  as  their difference  in  magnitude  should stay  the  same
\citep[e.g.][]{2018MNRAS.481L.130G}.    Therefore,   we    assume   by
hypothesis  that the  double peak  is  caused by  RC overdensities  at
different distances along the line of sight. Fig.~\ref{fig:view} shows
the position of the RC peaks along the three windows overplotted with a
schematic map  of the  MW adapted from  \cite{2009PASP..121..213C}. If
our assumption is correct, stars  belonging to these peaks are located
at different  positions along the  Galactic disk and  therefore should
present different  bulk motions as  a consequence of  the differential
Galactic rotation.

The     VVV    InfraRed     Astrometric    Catalogue     \citep[VIRAC,
][]{2018MNRAS.474.1826S} is a proper  motion and parallax catalogue of
the VVV survey.  The catalogue  includes 119 million sources with high
quality  proper   motion  measurements,  of  which   47  million  have
uncertainties                  below                  1~mas\,yr$^{-1}$
\citep{2018MNRAS.474.1826S}. We  made use  of the  enhanced v2
  version of  VIRAC called  VIRAC2 (Smith  et al.,  in prep.),  with a
  relative  to  absolute  correction  for  the  PMs  using  Gaia  EDR3
  \citep{2021A&A...649A...1G}. 
  
\begin{figure}
\includegraphics[scale={1.1}]{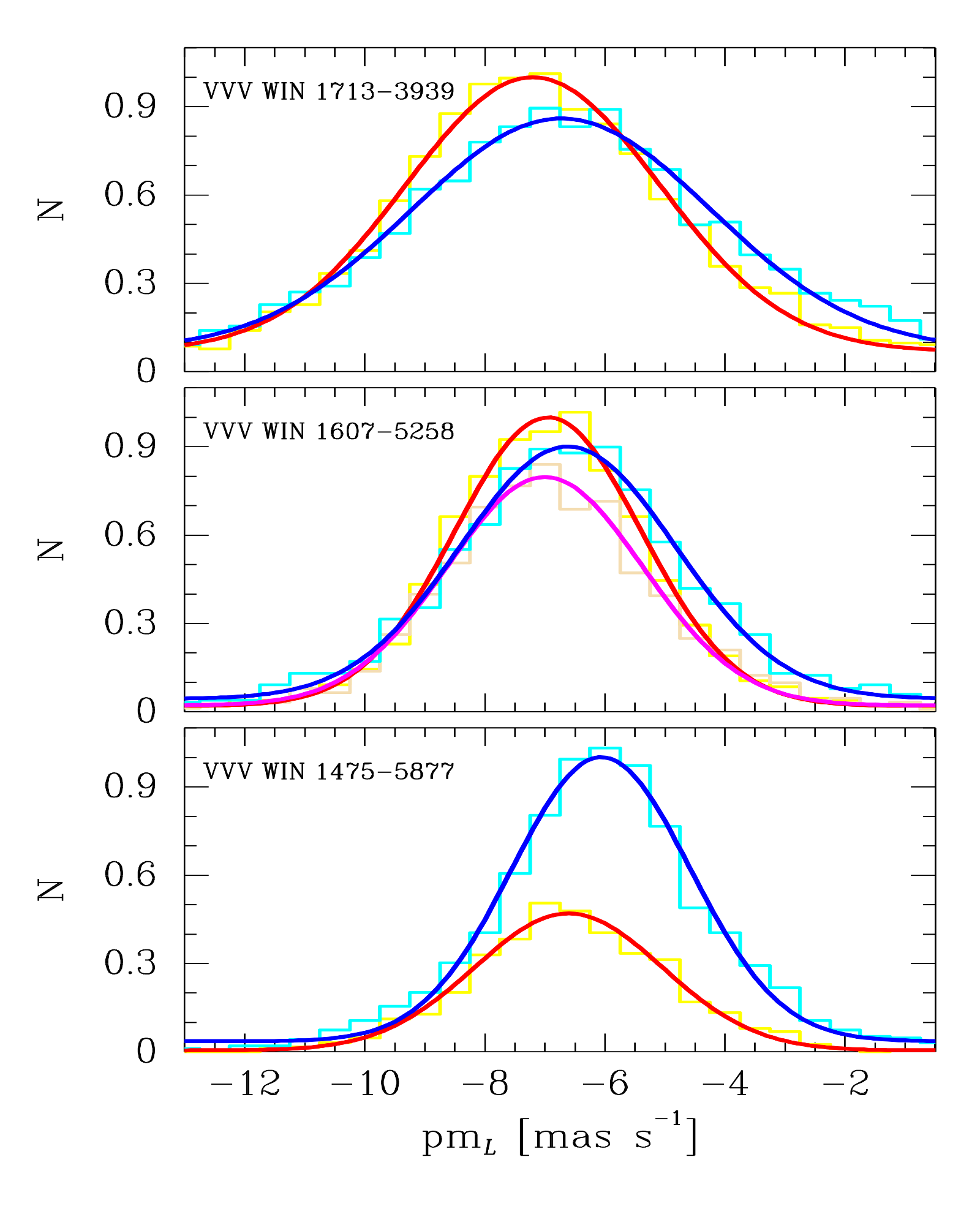}
\caption{Galactic  longitude component  of  the proper  motion
    ($\mu_{l}$) for  the RC  stars within the  three windows:  VVV WIN
    1713$-$3939 (top panel), VVV WIN  1607$-$5258 (middle) and VVV WIN
    1475$-$5877  (bottom).   The   color  code  is  the   same  as  in
    Figs.~\ref{fig:cmd}, \ref{fig:cmd2}  and \ref{fig:cmd3}:  RC stars
    belonging to  the closer peak are  shown in red while  those of the more distant peak are shown in blue. For  VVV WIN 1607$-$5258 (middle  panel) a
    third and intermediate peak appears in magenta.}
\label{fig:pm}
\end{figure}

Stars belonging  to each RC  were selected from the  CMDs (the
  diamond  shaped   boxes  on  the  CMDs   of  Figs.~\ref{fig:cmd}  to
  \ref{fig:cmd3})  and  then the  PMs  in  the equatorial  coordinates
  ($\mu_\alpha$,   $\mu_\delta$)    were   taken   from    VIRAC.    A
  transformation to  the Galactic  coordinates ($\mu_l$,  $\mu_b$) was
  performed using the method  described in \cite {2013arXiv1306.2945P}
  and then  a Gaussian fit was  applied to the PM  distribution to get
  the   mean    value   for   each   population    as   presented   in
  Fig.~\ref{fig:pm}).   Finally, we  converted  the  absolute PMs  to
physical velocities using the RCs  distances for each window. Although
our sample contains  stars that are not part of  the spiral arms, they
do not affect our results, due to their low quantity, when compared to
RC stars.  Table 1 summarizes these measurement values.

\begin{figure}
\hspace*{0.1cm}
\includegraphics[scale={0.425}]{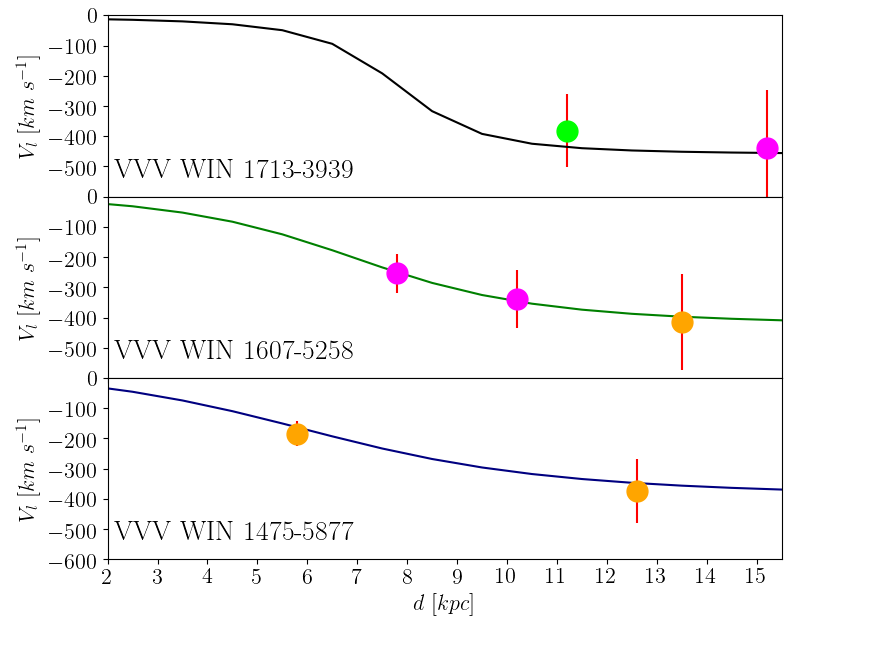}
\caption{Tangential   velocities   versus   distances   for   our   RC
  distributions along  the direction  of the three  windows considered
  here (located at $l=347.4$, $329.95$ and $318.65$~deg, respectively,
  from  top to  bottom),  compared  with a  rotation  curve model,  as
  described in  the text.  Data-points correspond  to the  observed RC
  peaks and follow the same color code used in Fig.~\ref {fig:view}.}
\label{fig:vel}    
\end{figure}

We  compare   our  results  with   a  simple  rotation   curve  model,
assuming  the distance  to  the Galactic  center  as 8.18  kpc
  \citep{2019A&A...625L..10G}, the velocity  of the solar neighborhood
  around the Galactic  center as $V_0 = 229$~ km\,s  $^{-1}$, and the
peculiar motion of the Sun with  respect to the Local Standard of Rest
(LSR)  as ($U_{\odot},  V_{\odot},  W_{\odot}$) =  (14.0, 12.0,  6.0)~
km\,s                $^{-1}$               \citep{2012MNRAS.427..274S,
  2014A&A...563A.128L}. Fig.~\ref{fig:vel}  presents the  $V_{l}$ $vs$
distance distribution  of our  data compared  with the  model, showing
excellent agreement.

\section{Mapping the spiral arms}

Our results demonstrate that both positions and velocities for the VVV
RC peaks are  consistent with expectations for the  spiral arms across
the disk.   In Fig.~\ref{fig:view} we note  that the RC peaks  in each
line of sight cross different arms, in good agreement with each other,
and    consistent   with    the   background    image   \citep[adapted
  from][]{2009PASP..121..213C} based  on radio,  IR, and  visible data
which is widely used as a current  view of our Galaxy. At the position
of ~VVV~WIN~1713$-$3939 the closer peak marks the end of the bar and/or
the   Perseus   arm   ($d=11.2$~kpc). Due to the  location  of 
~VVV WIN 1713-3939 it is  not  clear  whether we are observing stars at 
the beginning  of  the Perseus arm or at the end of the galactic bar.
The  first  peak in  VVV~WIN~1607$-$5258 and the more  distant peak in
VVV~WIN~1713$-$3939 mark the  position  of  the Norma arm ($d=9.0$ and
$14.7$~kpc)  while the two RC  peaks  in  VVV~WIN~1475$-$5877 and  the  
more  distant peak in VVV~WIN~1607$-$5258  mark the  position  of  the
Scutum-Centaurus arm ($d=6.8$, $11.8$ and $14.0$~kpc). VVV data saturate
for closer RC stars ($K_s < 12$~mag corresponds  to $d \lesssim 5$~kpc).
The  tangential  velocities  are also in good agreement with 
expectations for a  rotation  disk  at that given distance, confirming 
that they share the bulk rotation of the MW disk.

\cite{{2018A&A...618A.168R}} has  mapped the dust distribution  in the
Galactic disk  out to 7  kpc, using RCs  and giants stars  from APOGEE
DR14.   This is  the first  dust map  where arm  structures have  been
detected.   However  this  map  is  limited  to  the  foreground  disk
($d<7~kpc$).  This  work traces  the location of  the Perseus  arm and
found evidences  at the  position of ~Orion  Spur/local arm,  parts of
~Sagittarius and  especially the Scutum-Centaurus arm,  which is shown
in  Fig.~\ref{fig:view}  overlaid on  our  result.   While Gaia  alone
cannot  detect  clearly  the  spiral   arms  \citep[see  the  maps  of
][]{2019A&A...628A..94A},  the   work  of  ~\cite{2018MNRAS.481L..21P}
based on Gaia  DR2 data also showed evidences  of ~three overdensities
of upper  main sequence stars, that  correspond to Sagittarius-Carina,
Orion Spur/local  and Perseus  arms.  Both works  study the  first and
second  quadrant  (anticlockwise  in  relation   to  the  Sun  in  our
Fig.~\ref{fig:view}), while our complementary  observations are in the
fourth quadrant, at the far side of the disk.

\section{Conclusions}

We searched for new windows located at low Galactic latitudes ($|b|<1$
~deg)  in the  MW  plane, reporting  the discovery  of  two new  clear
near-IR windows: named VVV~WIN~1607$-$5258 and VVV~WIN~1475$-$5877. We
analyzed these  windows as  well as  the previously  discovered window
VVV~WIN~1713$-$3939.  The  CMD for each  window shows the  presence of
multiple RCs, with peaks  at different magnitudes, indicating
the presence  of structures along  the line  of sight. Our  work would
help to improve  future Milky Way models, as these  structures are not
present in  the current  models of synthetic  population for  the same
lines of sight along the Galactic disk.

Distances for  the RC are  consistent with the expected  positions for
the  spiral arms  at the  given lines  of sight  while the  tangential
velocities for the RC stars  increase with the distance, corroborating
our  interpretation of  structures at  different distances  across the
Galactic  disk. Tangential  velocities are  in good  agreement with  a
simple rotation disk  model. The signature of the Perseus  arm (or the
end of  the bar), the Norma  arm and Scutum-Centaurus arm  are mapped.
Our  results indicate  that the  distribution of  ~RC stars  along low
extinction windows are consistent both in distance and velocities with
expectations for the spiral arms in the MW disk.

\begin{acknowledgement}
We gratefully acknowledge  the use of data from the  ESO Public Survey
program  ID  179.B-2002  taken  with the  VISTA  telescope,  and  data
products  from the  Cambridge Astronomical  Survey Unit  (CASU).  R.K.
acknowledges support  from CNPq/Brazil.  R.K.S.   acknowledges support
from  CNPq/Brazil through  projects  308968/2016-6 and  421687/2016-9.
J.A.-G., acknowledges support from Fondecyt Regular 1201490 and from
ANID – Millennium Science Initiative Program – ICN12\_009 awarded to the
Millennium Institute of Astrophysics MAS. D.M.  gratefully  acknowledges
support  by the  ANID  BASAL  projects ACE210002 and  FB210003 and by 
Fondecyt Project No. 1220724,  and the Ministry  for   the  Economy,  
Development,  and   Tourism,  Programa Iniciativa Cient\'ifica Milenio
through grant IC120009, awarded to the Millennium Institute of
Astrophysics (MAS).
\end{acknowledgement}





\end{document}